\documentclass[twocolumn,epjc3]{svjour3}  
\smartqed  
\RequirePackage{graphicx}
\RequirePackage{amsmath}
\RequirePackage{amsfonts}
\RequirePackage{amssymb}
\RequirePackage{hyphenat}
\RequirePackage{url}
\RequirePackage{hyperref}
\usepackage{xcolor}
\journalname{Eur. Phys. J. A}
\begin{document}

\title{Re-investigation of heat capacity and paring phase transition in hot $^{93-98}$Mo nuclei}

\author{Le Thi Quynh Huong\thanksref{addr1,e1}
        \and
        Tran Dong Xuan\thanksref{addr2,addr3} 
        \and
        Nguyen Ngoc Anh\thanksref{addr4} 
        \and
        Nguyen Quang Hung\thanksref{addr2,addr3,e2}.
}

\thankstext{e1}{Corresponding author; e-mail: lethiquynhhuong@ukh.edu.vn}
\thankstext{e2}{Corresponding author: e-mail: nguyenquanghung5@duytan.edu.vn}


\institute{Department of Natural Science and Technology , Khanh Hoa University, Nha Trang City, Khanh Hoa Province, Vietnam \label{addr1}
           \and
           Institute of Fundamental and Applied Sciences, Duy Tan University, Ho Chi Minh City 700000, Vietnam \label{addr2}
           \and
           Faculty of Natural Sciences, Duy Tan University, Danang City 550000, Vietnam\label{addr3}
           \and
           Dalat Nuclear Research Institute, Vietnam Atomic Energy Institute, 01 Nguyen Tu Luc, Dalat City 670000, Vietnam\label{addr4}
}

\date{Received: date / Accepted: date}

\maketitle

\begin{abstract}
The empirical heat capacities of $^{93-98}$Mo nuclei are re-investigated by using the latest updated and recommended nuclear level density (NLD) data below the neutron binding energy $B_n$ combined with the back-shifted Fermi-gas (BSFG) model for the energy region above $B_n$. For the latter, the BSFG formula with energy-dependent level density parameter is used and the new parameterization has been carried out in order to obtain the best fit to the new NLD data in the whole data range. The results obtained show that the S-shaped heat capacity, a fingerprint of the pairing phase transition, is more pronounced in even $^{94,96,98}$Mo nuclei than that in odd $^{93,95,97}$Mo isotopes. This result is different with those obtained in two previous studies by R. Chankova et al., [Phys. Rev. C {\bf 73}, 034311 (2006)]  and K. Kaneko et al., [Phys. Rev. C {\bf 74}, 024325 (2006)], in which the old NLD data and the BSFG model with energy-independent level density parameter were used. Moreover, the present work suggests that the very strong S-shape observed in the heat capacities of both even and odd Molybdenum isotopes by K. Kaneko et al., [Phys. Rev. C {\bf 74}, 024325 (2006)] should be re-investigated. The present work also suggests that obtain the correct heat capacity and associated pairing phase transition in excited nuclei, one should use the correct NLD data and the best fitted BSFG NLD in the entire region where the experimental data are available.

\keywords{Nuclear level density \and Back-shifted Fermi-gas model \and Empirical heat capacity \and Hot $^{93-98}$Mo nuclei}
\end{abstract}

\section{Introduction}
\label{intro}
The nuclear level density (NLD), which was first introduced as the number of excited levels per unit of excitation energy \cite{Bethe1937}, is known to play a major role in studying not only nuclear structure and reactions, but also nuclear astrophysics \cite{Raus1,Raus2}. It also reflects the average properties of excited nuclei such as thermodynamic quantities (excitation energy, free energy, entropy, heat capacity, etc.) and/or quantum phase transition \cite{HungROPP}. In principle, the thermodynamic properties of a nucleus can be directly extracted from the experimental NLD data by using two statistical ensembles, namely the canonical ensemble (CE) and microcanonical ensemble (MCE) as atomic nucleus is a system with a fixed number of particles \cite{Suma07,Hung09,Hung17}. However, there are some ambiguities with the use of MCE since the experimental NLD is discrete and covers the excitation energy $E$ below the neutron binding energy $B_n$ ($\sim$ 8 MeV) only \cite{Oslo95,Oslo96,Oslo01,Oslo04,Oslo15}. For example, the nuclear temperature $T$ extracted from the experimental NLD data using the MCE can even become unphysically negative \cite{Mel01,Gut13}. Hence, the CE is often used to describe the nuclear thermodynamic properties instead of the MCE. Within the CE, the experimental NLD is extrapolated, by using the back-shifted Fermi-gas (BSFG) model \cite{Gilbert1965}, to cover the excitation energy above $B_n$ and up to 150 MeV. The experimental plus BSFG NLD is then used to construct the CE partition function $Z_{\rm CE}(T)$ based on which all the nuclear thermodynamic quantities are calculated \cite{Oslo99,Oslo00}.

Among the nuclear thermodynamic quantities extracted from the CE, the heat capacity, which characterizes the variation of $E$ with $T$, contains the information on the phase transition from a phase with strong pairing correlations to that with weak pairing ones via its S-shaped curve \cite{Egido2000,Liu2001,Schiller2001,Chankova2006,Kaneko2006,Dey17,Dey19}. However, the relation between this feature of heat capacity found in hot nuclei and the odd-even mass differences still remains disagreement. For example, the very unusually strong S-shaped heat capacities have been reported in a series of $^{93-98}$Mo nuclei in Ref. \cite{Kaneko2006}, which are quite different from those reported previously in Ref. \cite{Chankova2006} using the same NLD data of the same nuclei. Moreover, the NLD data of these nuclei have been recently re-analyzed and updated in Ref. \cite{exp13} based on a simultaneous analysis of the ($\gamma, n$) and ($n, \gamma$) reactions in which the NLD is one of the most important inputs. Hence, the heat capacities in these $^{93-98}$Mo should be re-investigated to have a precise description of the pairing phase transition in these isotopes. This is the goal of the present paper. 

\section{Formalism}
\label{sec:1}

\subsection{The back-shifted Fermi-gas model}
\label{sec:1.1}
The first NLD model, which was proposed long time ago by Bethe in 1937 \cite{Bethe1937}, is the Fermi-gas (FG) model. This model was derived based on an assumption that the single-particle states, from which the excited levels of the nucleus are constructed, are equally spaced. This simple assumption of the FG model is not enough to describe the NLD of excited nuclei in which several structures such as pairing correlation, shell effect, deformation, collective and rotational excitations, etc., are known to have significant contribution to the NLD. Hence, an extension of the FG model, called the back-shifted Fermi gas (BSFG) model, was proposed in Ref. \cite{Gilbert1965}, in which most of important structures of excited nuclei are taken into account. The BSFG formula has the form as
\begin{equation}
	\rho_{\rm BSFG}(E)  = \frac{\exp\left[2 \sqrt{a(E-E_1)}\right]}{12\sqrt2  \sigma a^{1/4}(E-E_1)^{5/4}}  ~,
	\label{rho} 
\end{equation}
where $a$ is the level density parameter; $E$ is the excitation energy; $E_1$ is the back-shifted energy, which is considered to be a free parameter determined by fitting to the data of each nucleus. The spin cut-off parameter $\sigma$ in Eq. (\ref{rho}) is given as
\begin{equation}
	\sigma^2 = 0.0888A^{2/3}\sqrt{a(E-E_1)} ~,
	\label{sigma} 
\end{equation}
with $A$ being the mass number.

The most important parameter in the BSFG formula (\ref{rho}) is the level density parameter $a$, which determines the slope of the NLD curve. In general, $a$ is assumed to be constant, whose values can be obtained from the systematic fitting to the experimental average level spacing $D_0$ data of a number of excited nuclei \cite{RIPL1,RIPL2}. However, microscopic NLD calculations have shown that the level density parameter $a$ should depend on the excitation energy \cite{Ig75,Ig79}. Moreover, the shell effect, which is large at low-excitation energy and damped at high-excitation energy, should be also taken into account \cite{Egidy05}. The energy-dependent level density parameter $a$ taking into account the damping of the shell effect can be described by the a phenomenological formula as follows \cite{Ig75}
\begin{align}
	a &\equiv a(Z,A,E) \nonumber \\ 
	  &= \widetilde{a}(A)\left\{1+ \frac{\delta W(Z,A)}{E-E_1}[1-e^{-\gamma(E-E_1)}]\right\} ~,
	\label{aLD} 
\end{align}
where $\widetilde{a}(A)$ is the asymptotic level density obtained when all the shell effects are damped, whereas $\delta W\{Z,A\}$ is the shell-correction energy defined as
\begin{equation}
	\delta W(Z,A) = M_{\rm exp} - M_{\rm LD} ~,
	\label{shellcorr} 
\end{equation}
with $M_{\rm exp}$ and $M_{\rm LD}$ being the experimental mass and the mass calculated by using the macroscopic liquid-drop formula, respectively. The damping parameter $\gamma$ in Eq. (\ref{aLD}), which determines how rapidly $a(Z, A,E)$ approaches $\widetilde{a}(A)$, is given as
\begin{equation}
	\gamma = \frac{\gamma_0}{A^\frac{1}{3}} ~,
	\label{gamma}
\end{equation}
where $\gamma_0$ is the global parameter obtained from the fittings to the experimental data over a whole range of mass numbers.

There is another version of the BSFG formula, which has been used in the normalization procedure of the experimental NLD within the Oslo technique (see e.g., Refs. \cite{Schiller2001,Chankova2006,Kaneko2006}). In this version, the level density parameter is assumed to be energy-independent. Therefore, to have the best fit to the experimental NLD data, an additional fitting parameter $\eta$ is used, namely 
\begin{equation}
	\rho_{\rm BSFG}(E)  = \eta\frac{\exp\left[2 \sqrt{a(E-E_1)}\right]}{12\sqrt2  \sigma a^{1/4}(E-E_1)^{5/4}}  ~,
	\label{rho1} 
\end{equation}
where $\sigma$ is the same as Eq. (\ref{sigma}); $a = 0.21A^{0.87}$; and $E_1 = C_1+E_{\rm pair}$ with $C_1 = -6.6A^{-0.32}$ and $E_{\rm pair}$ being the pairing energy. The parameter $\eta$ is adjusted in order to reproduce the experimental $D_0$ data \cite{Kaneko2006}. This version of BSFG formula is only used to connect the NLD data in the intermediate energy region from $\sim 4$ to ($B_n - 1$) MeV to the NLD data at $E=B_n$. Therefore, for the overall description of the NLD data in the whole energy region, the use of formula (\ref{rho}) (with the energy-dependent $a$) should be always better than that of Eq. (\ref{rho1}) (with the energy-independent $a$). For instance, the BSFG with energy-dependent $a$ has been widely used in the normalization procedure within the Oslo method since 2013 (see e.g., Refs. \cite{LarsenPRC87,TornyiPRC89}), whereas the BSFG with energy-independent $a$ was only used before that (see e.g., Refs. \cite{Schiller2001,Chankova2006}).

\subsection{Thermodynamic quantities}
\label{sec:1.2}

Having a highly dense state density, even at low-excitation energy, an atomic nucleus can be well described by a statistical ensemble \cite{Suma07,Hung09}. Hence, the NLD has a direct relation to the thermodynamic partition function, which characterizes the statistical properties of a system in thermodynamic equilibrium. As discussed in the Introduction, the canonical ensemble (CE) is often used to describe the nuclear thermodynamic properties. 

Within this ensemble, the CE partition function $Z(T)$ can be calculated from the NLD making use of the inverse Laplace transform of the NLD, namely \cite{BM}
\begin{equation}
Z(T)= \sum_{E_i=0}^{\infty} \rho(E_i)e^{-E_i/T}\delta E_i ~,
\label{ZT}
\end{equation}
where $\rho(E_i)$ is the total NLD at a given excitation energy $E_i$; $\delta E_i$ is the energy bin; and $T$ is nuclear temperature. Based on the CE partition function (\ref{ZT}), we can easily calculate all the thermodynamic quantities such as free energy $F=-T{\rm ln}Z$, entropy $S=-\partial F/\partial T$, total energy $\overline{E}=F+TS$, and heat capacity $C=\partial\overline{E}/\partial T$.

It is worthwhile to mention that the summation in Eq. (\ref{ZT}) should be taken from $E_i = 0$ to $E_i \rightarrow \infty$, that is, the NLD $\rho(E_i)$ must be extended up to very high-excitation energy. However, the experimental NLD data are usually limited below the neutron binding energy $B_n$. Hence, the BSFG model (\ref{rho}) is used to extend the NLD to above $B_n$ and up to 120 MeV. This choice of maximum $E_i$ is enough for the study of nuclear thermodynamics within a temperature range from 0 to 2 MeV \cite{HungPRC81,HungPRC82}. Consequently, the summation in Eq. (\ref{ZT}) is split into two parts, namely 
\begin{align}
	Z(T)= & \sum_{E_i=0}^{E_i < B_n} \rho_{\rm exp}(E_i)e^{-E_i/T}\delta E_i \nonumber 
	\\
		  & + \sum_{E_i=B_n}^{E_i = 120~{\rm MeV}} \rho_{\rm BSFG}(E_i)e^{-E_i/T}\delta E_i ~,
	\label{ZT1}
\end{align}
where the first part is related to the experimental NLD data ($\rho_{\rm exp}$), while the second part is associated to the BSFG NLD ($\rho_{\rm BSFG}$). The thermodynamic quantities obtained from the partition function (\ref{ZT1}) are often called the experimental or empirical ones, which are used to compare with the theoretical calculations (see e.g., Refs. \cite{Schiller2001,Chankova2006,Kaneko2006,HungPRC81,HungPRC82}).

\section{Numerical results and discussion}
\label{sec:2}

Six medium-mass $^{93-98}$Mo nuclei, whose experimental NLD data are available in Refs. \cite{Chankova2006,exp13,exp03}, have been selected in the present work. Regarding the experimental NLD data, the first datasets for two $^{96,97}$Mo nuclei were available in 2003 using the ($^3$He, $^3$He$^{'}$) reactions \cite{exp03}. The later experiments using both ($^3$He, $^3$He$^{'}$) and ($^3$He, $\alpha$) reactions in 2006 provided another datasets for all $^{93-98}$Mo nuclei \cite{Chankova2006}. In the latest experimental study in 2013, the previous NLD data of $^{94-98}$Mo have been re-analyzed by self-consistently combining the photoabsorption cross-section data obtained within the ($^3$He, $^3$He$^{'}$) and ($^3$He, $\alpha$) reactions with those taken from the ($n, \gamma$) and ($\gamma, n$) experiments \cite{exp13}. In the latest experiment, an adopted range for the $D_0$ values of $^{94-98}$Mo, based on which the experimental NLD data were normalized, has been proposed. This range includes the minimum $D_0^{\rm min}$, maximum $D_0^{\rm max}$, and recommended $D_0^{\rm rec}$ values \cite{exp13} and the new NLD datasets normalized using those $D_0$ values are provided and accessible via Ref. \cite{uio}. In the present work, the recommended $D_0^{\rm rec}$ and its associated NLD data have been used. 

\begin{table}[h]
	\caption{Parameters of the BSFG formula (\ref{rho1}) used in Refs. \cite{Chankova2006,Kaneko2006}}
	\label{tab1}       
	
	\begin{tabular}{p{3.5em}p{3em}p{4em}p{3em}p{2em}p{2.5em}}
		\hline
		Nucleus & $E_{\rm pair}$ (MeV) & $a$ (MeV$^{-1}$) & $C_1$ (MeV) & $\eta$ & $D_0$ (eV) \\
		\hline
		$^{93}$Mo & 0.899 & 10.83 & -1.547 & 0.08 & 2700 \\
		$^{94}$Mo & 2.027 & 10.93 & -1.542 & 0.25 & $-$ \\
		$^{95}$Mo & 1.047 & 11.03 & -1.537 & 0.34 & 1320 \\
		$^{96}$Mo & 2.138 & 11.13 & -1.531 & 0.46 & 105 \\
		$^{97}$Mo & 0.995 & 11.23 & -1.526 & 0.65 & 1050 \\
		$^{98}$Mo & 2.080 & 11.33 & -1.521 & 0.87 & 75 \\
		\hline
	\end{tabular}
\end{table}

\begin{table}[h]
	\caption{Parameters of the BSFG formula (\ref{rho}) used in this work in comparison with those taken from RIPL-3 database \cite{RIPL3}}
	\label{tab2}

	\begin{tabular}{p{3.5em}p{3em}p{3.5em}p{3.5em}p{3.5em}p{2.5em}}
		\hline
		\multicolumn{6}{c}{RIPL-3}\\
		\hline
		Nucleus & $\delta W$ (MeV) & $\gamma$ (MeV$^{-1}$) & $\tilde{a}$ (MeV$^{-1}$) & $E_1$ (MeV) & $D_0$ (eV) \\
		\hline
		$^{93}$Mo & -1.843 & 0.091 & 10.370 & -0.095 & 2700 \\
		$^{94}$Mo & -0.473 & $-$ & $-$ & $-$ & $-$ \\
		$^{95}$Mo & 0.097 & 0.090 & 10.892 & 0.022 & 1320 \\
		$^{96}$Mo & 1.025 & 0.090 & 10.709 & 1.141 & 105 \\
		$^{97}$Mo & 1.700 & 0.089 & 10.745 & -0.131 &  1050 \\
		$^{98}$Mo & 2.453 & 0.089 & 10.811 & 1.024 & 75 \\
		\hline
	\end{tabular}
	
	\begin{tabular}{p{3.5em}p{3em}p{3.5em}p{3.5em}p{3.5em}p{2.5em}}
		\multicolumn{6}{c}{This work}\\
		\hline
		Nucleus & $\delta W$ (MeV) & $\gamma$ (MeV$^{-1}$) & $\tilde{a}$ (MeV$^{-1}$) & $E_1$ (MeV) & $D_0$ (eV) \\
		\hline
		$^{93}$Mo & -1.843 & 0.101 & 10.070 & 0.055 & 2700 \\
		$^{94}$Mo & -0.473 & 0.081 & 10.495 & 0.801 & 81 \\
		$^{95}$Mo & 0.097 & 0.110 & 11.152 & 0.217 & 831 \\
		$^{96}$Mo & 1.025 & 0.080 & 10.509 & 0.879 & 66 \\
		$^{97}$Mo & 1.700 & 0.089 & 10.545 & -0.226 & 661 \\
		$^{98}$Mo & 2.453 & 0.036 & 11.071 & 0.764 & 47 \\
		\hline
	\end{tabular}
\end{table}

\begin{figure}[h]
	\includegraphics[scale=0.38]{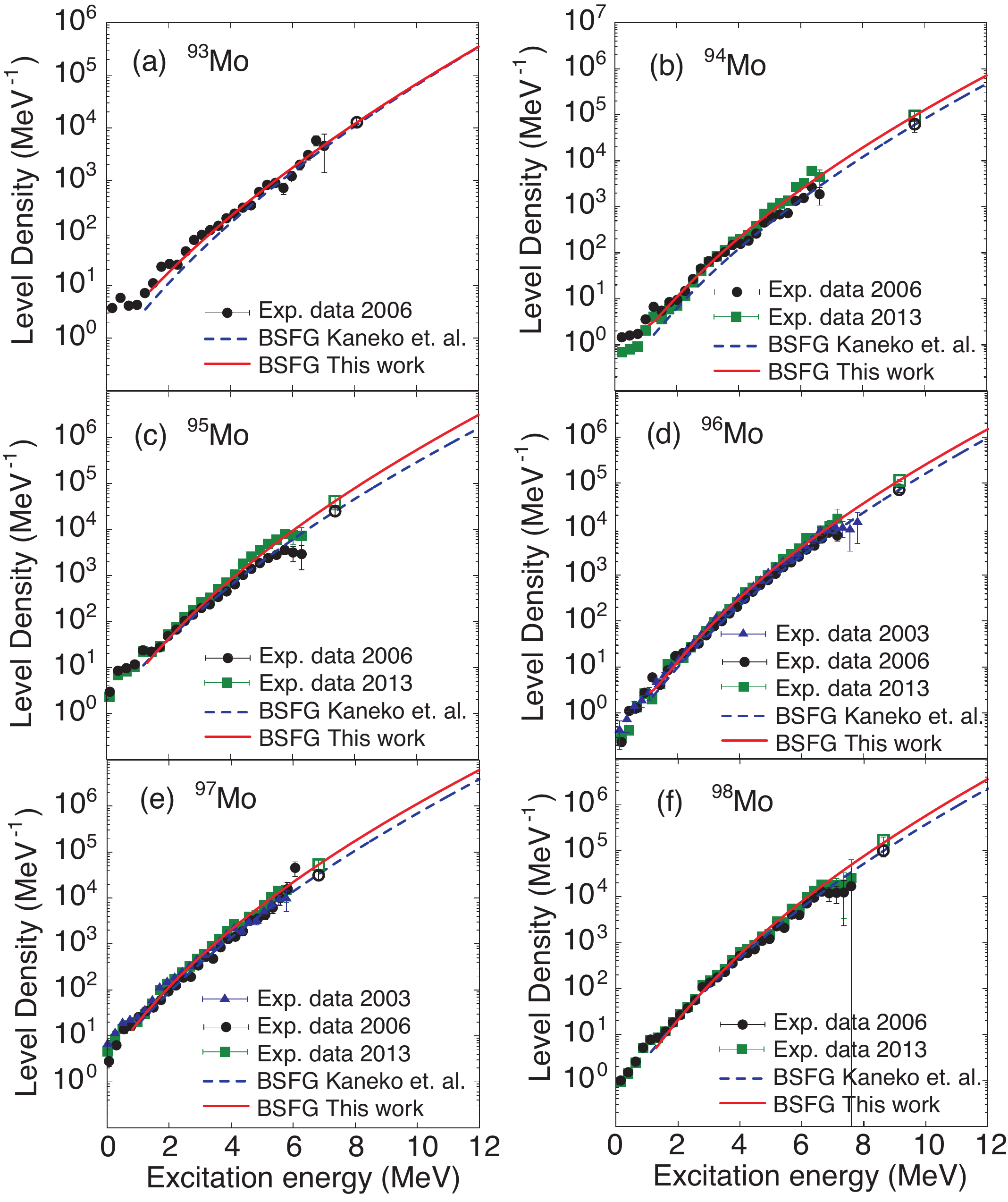}
	\caption{Experimental and BSFG NLDs as functions of excitation energy for $^{93-98}$Mo. The filled triangles and circles (with error bars) are the NLD data taken from Refs. \cite{exp03} and \cite{Chankova2006}, respectively. The filled squares (with error bars) are the recommended NLD data taken from Refs. \cite{exp13,uio}. The open circles are the NLD data at the neutron binding energy $B_n$ taken from Fig. 4 of Ref. \cite{Chankova2006}. The open squares are the NLD data at the neutron binding energy $B_n$ calculated from the recommended $D_0$ data in Table I of Ref. \cite{exp13} using the BSFG parameters taken from Table \ref{tab2}. The dashed and solid lines are the NLDs obtained by using the BSFG formulae (\ref{rho1}) and (\ref{rho}) with parameters listed in Tables \ref{tab1} and \ref{tab2}, respectively.}
	\label{fig1}
\end{figure}

\begin{figure}[h]
	\includegraphics[scale=0.38]{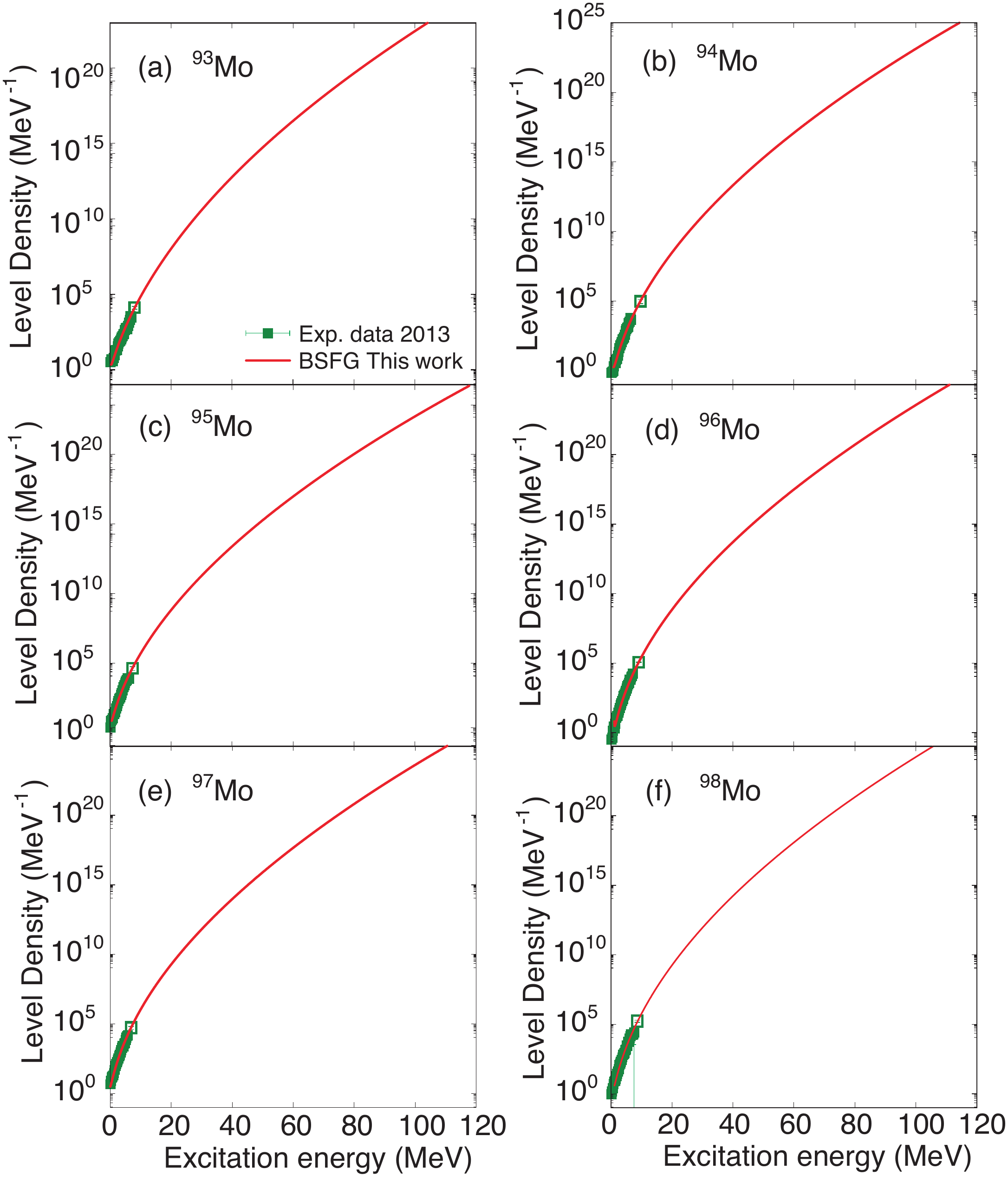}
	\caption{Experimental and BSFG NLDs as functions of excitation energy up to $120$ MeV for $^{93-98}$Mo. Experimental data (squares) are taken from the recommended values in Refs. \cite{exp13,uio}, whereas the BSFG NLDs (solid lines) are calculated using Eq. \ref{rho} with parameters listed in Table \ref{tab1}.}
	\label{fig2}
\end{figure}

All the available NLD datasets of $^{93-98}$Mo are plotted in Fig. \ref{fig1} along with those obtained from the BSFG model using two parameterizations in Eqs. (\ref{rho}) and (\ref{rho1}). The parameters of the BSFG formula (\ref{rho1}) together with the $D_0$ values are taken from Refs. \cite{Chankova2006,Kaneko2006} and listed in Table \ref{tab1}. Since the new NLD and $D_0$ data for $^{94-98}$Mo are recommended, the old BSFG parametrization in Table \ref{tab1} is no longer valid. This can be clearly seen in Figs. \ref{fig1}[(b)-(f)] that the BSFG NLDs with parameters in Table \ref{tab1} (dashed lines) agree with the NLD datasets in 2003 (triangles) and 2006 (circles) only. Hence, the new BSFG parameterization must be performed in order to obtain the best fit to the newly recommended NLD datasets (squares in Fig. \ref{fig1}). In this work, we used the BSFG formula (\ref{rho}) and slightly re-adjusted its parameters to obtain the best fit to the new NLD datasets. These new BSFG parameters and $D_0^{\rm rec}$ values are listed in Table \ref{tab2} in comparison with those taken from RIPL-3 database \cite{RIPL3}. It is clearly seen in Figs. \ref{fig1}[(b)-(f)] that the BSFG NLDs obtained using these new parameters (solid lines) fit very well the new datasets (squares), except for $^{93}$Mo. For $^{93}$Mo, there exists only one dataset in 2006. Moreover, the BSFG with parametrization in Table \ref{tab1} even underestimates the experimental data below about 4 MeV. Therefore, we have to readjust the BSFG parameters to obtain the best fit to the experimental data in the entire data range (from about $1-8$ MeV) (see Fig. \ref{fig1}(a) and Table \ref{tab1}). In Fig. \ref{fig2}, we plot the extension of the BSFG NLDs with new parameters up to about 120 MeV. Obviously, the new BSFG NLDs reproduce very well the slope of the experimental data.
\begin{figure}[h]
	\includegraphics[scale=0.44]{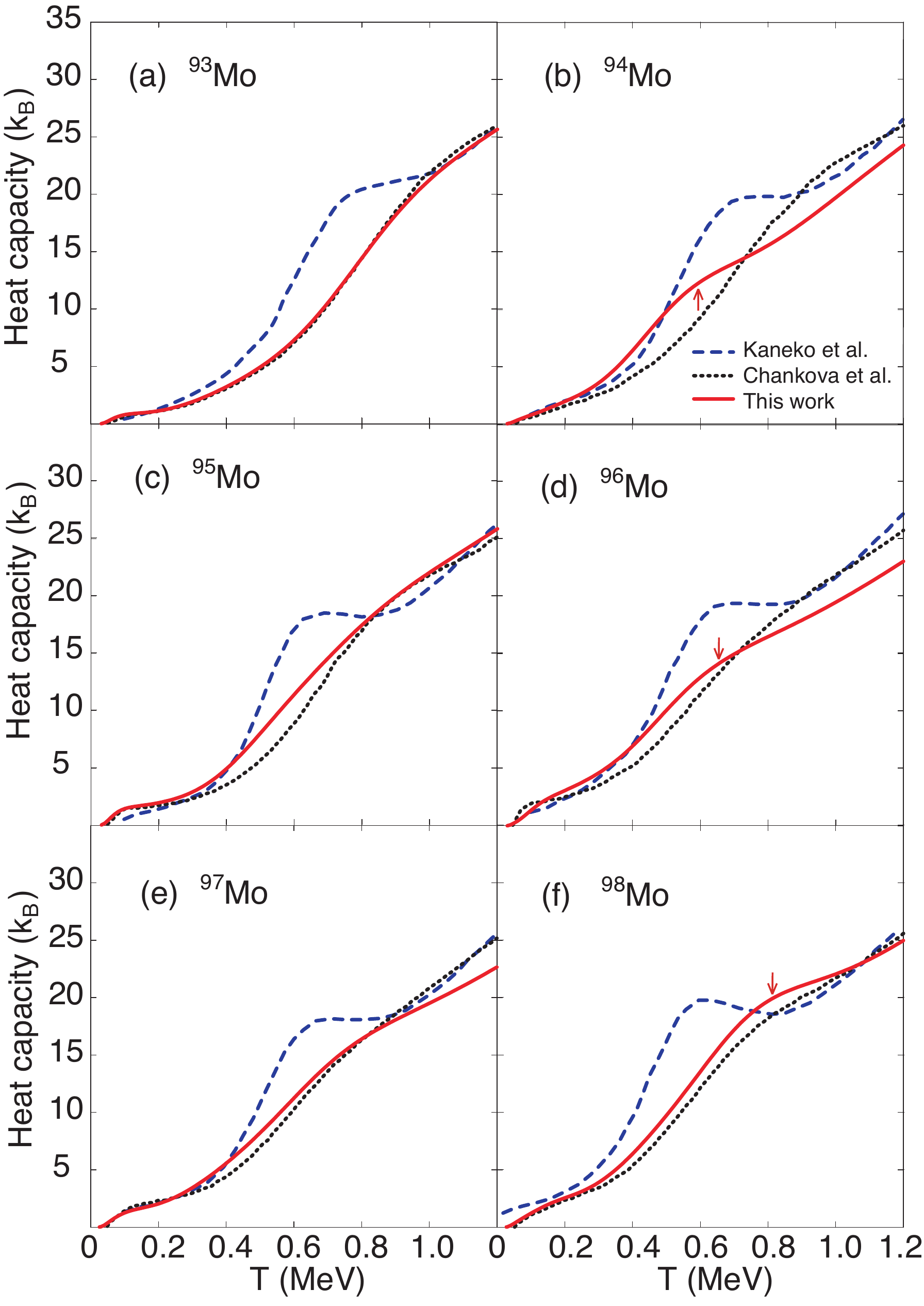}
	\caption{Empirical heat capacities as functions of temperature $T$ for $^{93-98}$Mo. The dashed lines present the exp+FG heat capacities taken from Fig. 3 of Ref. \cite{Kaneko2006}, whereas the dotted lines denote those taken from Fig. 10 of Ref. \cite{Chankova2006}. The solid lines stand for the results obtained within the present work by using the new exp+BSFG NLDs in Fig. \ref{fig2}. The red arrows denote the region where the S-shaped heat capacity is observed within the present work.}
	\label{fig3}
\end{figure}

The empirical heat capacities of $^{93-98}$Mo calculated using the partition function (\ref{ZT1}) and the exp+BSFG NLDs in Fig. \ref{fig2} are plotted in Fig. \ref{fig3} versus those taken from previous studies in Refs. \cite{Chankova2006,Kaneko2006}. In this figure, the heat capacities taken from Fig. 3 of Ref. \cite{Kaneko2006} (dashed lines) show a very strong S-shape, whereas those taken from Fig. 10 of Ref. \cite{Chankova2006} exhibit mostly none S-shape. This distinct difference between these two heat capacities is quite strange because they are calculated using the same exp+BSFG NLD data. Indeed, we have tried to re-calculate the heat capacities using exactly the same experimental NLD data (circles in Fig. \ref{fig1}) plus the same BSFG NLDs (dashed lines in Fig. \ref{fig1}). The results obtained are almost coincided with those taken from Fig. 10 of \cite{Chankova2006}. We have also performed another test that we take the solid lines (exp+FG) in Fig. 2 of Ref. \cite{Kaneko2006} and re-calculate the heat capacities. However, we cannot reproduce the exp+FG heat capacities in Ref. \cite{Kaneko2006}. This examination suggests that there might be an error in the numerical calculation of heat capacities in Ref. \cite{Kaneko2006}. By using the new NLD and BSFG data in Fig. \ref{fig2}, the obtained heat capacities (solid lines in Fig. \ref{fig3}) show a qualitative difference with those taken from two previous studies, namely the S-shape is observed in even-even $^{94,96,98}$Mo nuclei (see the heat capacities around the region indicated by the arrows in Fig. \ref{fig3}), whereas it is invisible in odd-A $^{93,95,97}$Mo isotopes. In addition, the new S-shape observed in even-even nuclei is stronger than that seen in Ref. \cite{Chankova2006} but weaker than that predicted in Ref. \cite{Kaneko2006}. This result is reasonable because of two reasons. The first reason is that the even-even nuclei often exhibit stronger pairing correlations than the odd ones. The second reason is that the even $^{94,96,98}$Mo nuclei are deformed in its ground state because their experimental quadrupole deformation parameters $\beta_2$ are 0.1511 $\pm$ 0.0016, 0.1715 $\pm$ 0.0014, and 0.1691 $\pm$ 0.0017, respectively \cite{Nudat2.8}. It is known that deformed nuclei always exhibit weaker pairing correlations than spherical ones \cite{Dey19}. This result also suggests that the strong S-shaped heat capacities obtained for odd $^{93,95,97}$Mo nuclei in Ref. \cite{Kaneko2006} should be re-investigated. In other words, the newly recommended NLD data in Ref. \cite{exp13} reflect the correct behavior of nuclear heat capacities in even and odd Molybdenum isotopes and should be used later on for the comparison with the theoretical calculations.

\section{Conclusions}

The present work re-investigates the empirical heat capacities of $^{93-98}$Mo nuclei extracted from the experimental NLDs below the neutron binding energy $B_n$ combined with the BSFG model for the energy above $B_n$ and up to about 120 MeV. For this purpose, the newly updated and recommended NLD data in 2013 have been used and combined with the BSFG NLD with energy-dependent level density parameter and new parameterization. The results obtained indicate that the heat capacities of even $^{94,96,98}$Mo nuclei show a more pronounced S-shape than those in odd $^{93,95,97}$Mo isotopes. This finding is qualitatively different with those obtained in two previous studies in Refs. \cite{Chankova2006,Kaneko2006}, in which the old NLD data and the BSFG NLD with energy-independent level density parameter are used. In particular, the present work suggests that the strong S-shape observed in the heat capacities of both even and odd Molybdenum isotopes in Ref. \cite{Kaneko2006} might be caused by an error in the numerical calculations, which should be re-investigated. Finally, to obtain the correct behavior of nuclear heat capacity and associated pairing phase transition, one should use the correct NLD data and the best fitted BSFG NLD in the entire region where the experimental data are available. 

\begin{acknowledgements}
The authors wish to thank University of Khanh Hoa for supporting through the research project No. KHTN-20.01. This work is funded by the National Foundation for Science and Technology Development (NAFOSTED) of Vietnam under Grant No. 103.04-2019.371.
\end{acknowledgements}

\end{document}